\documentclass[letter]{aa} 

\usepackage{graphicx}
\usepackage{txfonts}
\usepackage{hyperref}
\usepackage{color}
\usepackage{soul}   
\begin{document} 

   \title{Stellar density profile and mass of the Milky Way Bulge from VVV data\thanks{Based on observations 
taken within the ESO/VISTA Public Survey VVV under the programme ID 179.B-2002 (PI: Minniti)}}

   \author{E. Valenti,
          \inst{1}
          M. Zoccali,
          \inst{2,3}
         O. A. Gonzalez,
          \inst{4,5}
          D. Minniti,
          \inst{3,6,7}
         J. Alonso-Garc\'{i}a,
          \inst{8,3}
          E. Marchetti,
          \inst{1}
          M. Hempel,
          \inst{2}
         A. Renzini,
          \inst{9}
          M. Rejkuba,
          \inst{1,10}
          }

\institute{European Southern Observatory, Karl Schwarzschild\--Stra\ss e 2, D\--85748 Garching 
bei M\"{u}nchen, Germany. \\
              \email{evalenti@eso.org}
\and
Instituto de Astrofi\'{i}sica, Pontificia Universidad Cat\'{o}lica de Chile, Av. Vicu\~{n}a Mackenna 4860, Santiago , Chile.
\and
Millennium Institute of Astrophysics, Av. Vicu\~{n}a Mackenna 4860, 782-0436 Macul, Santiago, Chile.
\and
European Southern Observatory, A. de Cordova 3107, Casilla 19001, Santiago 19, Chile
\and
Institute for Astronomy, University of Edinburgh, Royal Observatory, Edinburgh EH9 3HJ
\and
Departamento de Ciencias Fis\'{i}cas, Universidad Andr\'{e}s Bello, Rep\'{u}blica 220, Santiago, Chile
\and
Vatican Observatory, V00120 Vatican City State, Italy
\and
{Unidad de Astronom\'\i a, Fac. de Ciencias B\'asicas, Universidad de Antofagasta, Antofagasta, Chile}
\and
Istituto Nazionale di Astrofisica, Osservatorio Astronomico di Padova,  Italy
\and
Excellence Cluster Universe, Boltzmannstr. 2, 85748 Garching, Germany
             }

   \date{}

 
  \abstract {We  present the first stellar density profile of the Milky Way bulge reaching latitude  $b=0^\circ$. 
 It is derived by counting red clump stars within the colour\--magnitude diagram constructed with the
    new PSF-fitting  photometry from  VISTA Variables in the V\'\i a  L\'actea (VVV) survey data. The new 
    stellar density map  covers the area between $|l|\leq 10^\circ$  and $|b|\leq 4.5^\circ$ with
    unprecedented accuracy, allowing to establish a direct  link between the stellar kinematics from
    the Giraffe  Inner Bulge  Spectroscopic Survey  (GIBS) and  the stellar  mass density  distribution. In
    particular, the location of the central velocity dispersion peak from GIBS matches a high overdensity in
    the VVV star count map.  By scaling the total luminosity function (LF) obtained from  all VVV fields 
    to the LF from Zoccali et al.(2003), we obtain the first fully empirical estimate of the mass in  stars and 
    remnants of the Galactic bulge. 
    The Milky Way bulge stellar mass within  ($|b|<9.5^\circ$, $|l|<10^\circ$) is $2.0\pm0.3\times 10^{10}M_{\odot}$.}
  \keywords{Galaxy: structure -- Galaxy: Bulge } 
  \authorrunning{Valenti et al.}  
  \maketitle
%

\section{Introduction}

Core helium\--burning  red clump  (RC) stars  are useful distance  indicators,
because their  magnitude varies slowly and smoothly with age  and metallicity, and is very well  predicted by
stellar evolution models \citep[e.g.,][]{salaris02}. The RC is also a bright and distinct feature in the 
colour\--magnitude diagram (CMD) of intermediate to old stellar populations, hence easy to identify. 
This makes it a powerful tracer of the Milky Way (MW) bulge structure.  In particular, RC
stars have been extensively  used to trace the bar  position, orientation and scale
parameters \citep[see the review of][for a complete  list of references]{oscarrev15}, as well as to
uncover the presence of a peanut (X-shape) in the outer bulge \citep{mczoc10,nataf10,saito11}.

The infrared  photometry from the VISTA  Variables in the V\'\i  a L\'actea (VVV) ESO  public survey
\citep{minniti_vvv} has been used by \citet{wegg13} to constrain the first 3\--D density model of
the  bulge. Because  the DR1 catalogue \citep{vvv_dr1}  used in  that study  is based  on aperture
photometry, which suffers strongly from incompleteness due to high level of crowding in  the region  $|b|\le0.5^\circ$, their  model was
rather poorly constrained close to the plane.

Here we present the first stellar density map reaching all the way to the centre of the MW bulge using RC star counts from new VVV catalogues based on PSF-fitting photometry.  We compare directly  the stellar  density with  kinematics, and  derive  the first fully  empirical estimate of  the stellar mass of the bulge.

\section{Observation and data reduction}

Our data are part of the VVV survey that covers contiguously $\sim$180 sq.\ deg of MW bulge between $|l|\leq10^{\circ}$ and $|b|\leq4.5^{\circ}$.  A detailed description of the global survey, and of the multi\--band observations completed within the first two years, which are used in this study, can be found  in \citet{minniti_vvv} and \citet{vvv_dr1}, respectively.

Unlike most previous studies based on DR1{\footnote { Catalogs obtained from aperture photometry on individual tiles.}} VVV data, here we retrieved the  {\it stacked pawprint} images from
the CASU pipeline{\footnote {http://casu.ast.cam.ac.uk/}}, and following \citet{alonsoG15}, we performed
PSF\--fitting photometry with the DoPHOT code \citep{dophot,javier_dophot} on each of the 16 chips in the images, both in $J$ and $K_s$.  
The photometric catalogues of the individual chips were calibrated and astrometrized to the VISTA system. 
The six pawprint catalogues covering a VISTA tile were combined into tile catalogues and magnitudes of stars detected multiple times were weight-averaged.

\begin{figure}
\centering
\includegraphics[width=9.0cm,angle=0]{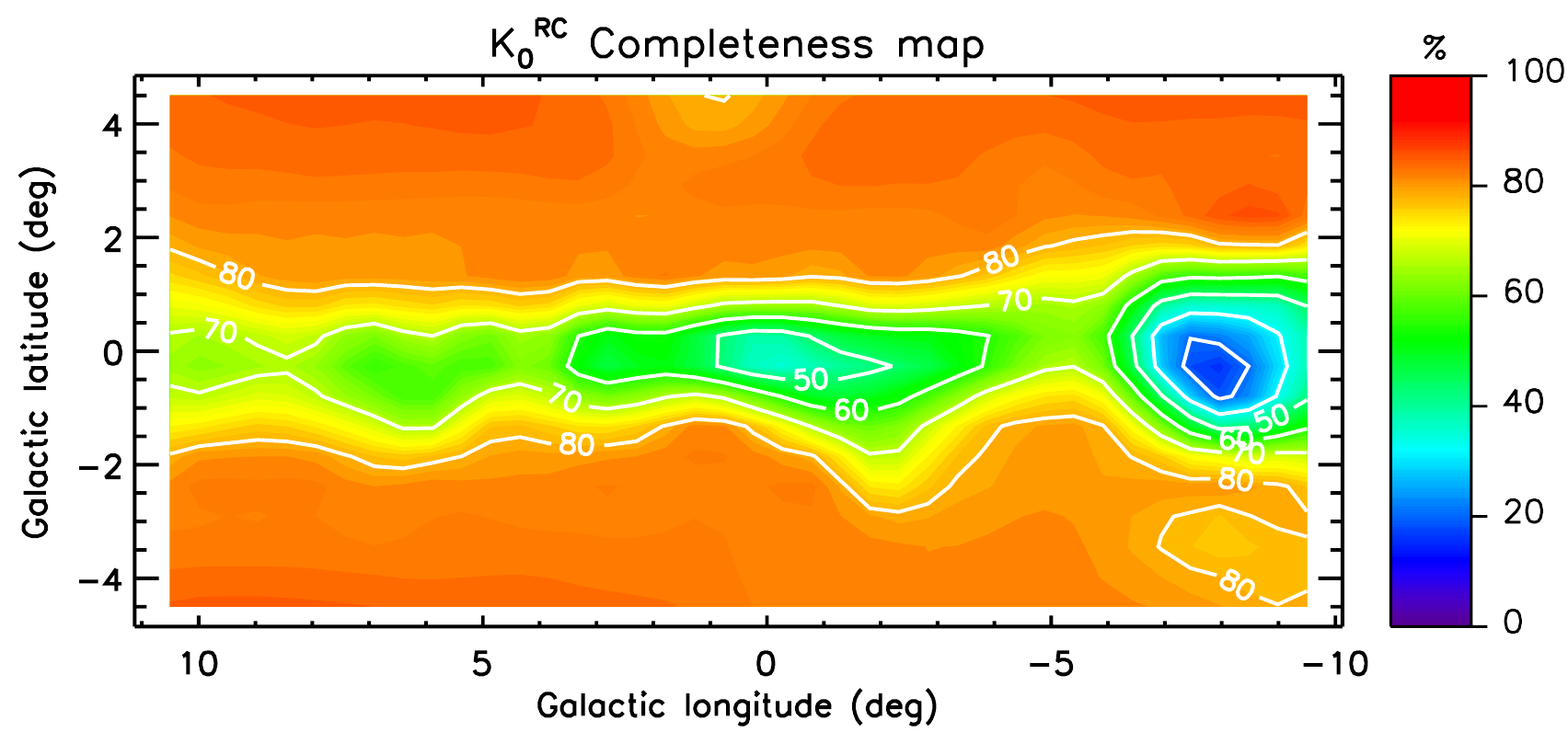}
\caption{Completeness map of the RC peak magnitude ($K_0^{RC}$) across the bulge. 
\label{compl}}
\end{figure}

The photometric  completeness was  estimated via artificial star experiments. 
For each tile, stars were randomly added only to chip\#12, under the assumption that the stellar density in this chip is representative of the whole tile.  
A more  detailed description of the  PSF photometry and completeness experiments for the  whole VVV survey area is the  subject of a forthcoming paper (Alonso\--Garcia  et al.  2015, {\it in prep.}).   
DoPHOT allowed us to obtain photometry with a  typical error $<0.05$ down to $J$$\sim$18 and $K_s$$\sim$17 in most of the bulge area.  
As expected, in the faint ($K_s$$>17$) magnitude regime the completeness of the catalogues can vary significantly across the bulge area, being lower for tiles closer to the plane, where both crowding and extinction are more severe.   
However, in the magnitude range of  interest for the present work, i.e.,  $K_0\leq14$, the derived catalogues are more than $70\%$ complete with the only exception of a few fields (see Fig.\ref{compl}).

\section{Analysis and results}
\label{sec:results}

Our aim  is to  investigate in  detail the  3\--dimensional structure and  symmetry of  the inner  bulge, by tracing the stellar density from RC  star counts. 
In order to build  a RC\--based stellar density  map we needed to address two main issues related to the  observing strategy adopted by VVV to survey the bulge area.  
The first is that there  is a non negligible overlap ($\sim$10$\%$) between adjacent tiles.
We therefore cross-correlated adjacent tiles eliminating double detections. 
The second issue is that the centres of the tiles are not symmetric with respect to the plane ($b=0^\circ$) nor to the bulge projected minor axis ($l=0^\circ$).  
Hence we resampled the bulge area analysed here with a new uniform grid of tiles, hereafter called {\it fields}, each with a $1^\circ \times 1^\circ$ area centred at  integer values of both latitude and longitude  between $-9.5^\circ\leq l \leq 10.5^\circ$ and $|b|  \leq  4.5^\circ$.   
This  allows  us  to  compare directly and consistently  fields  located symmetrically about  the bulge projected axes  without the need  for interpolations on stellar  density or completeness.

\begin{figure}
\centering
\includegraphics[width=5.5cm]{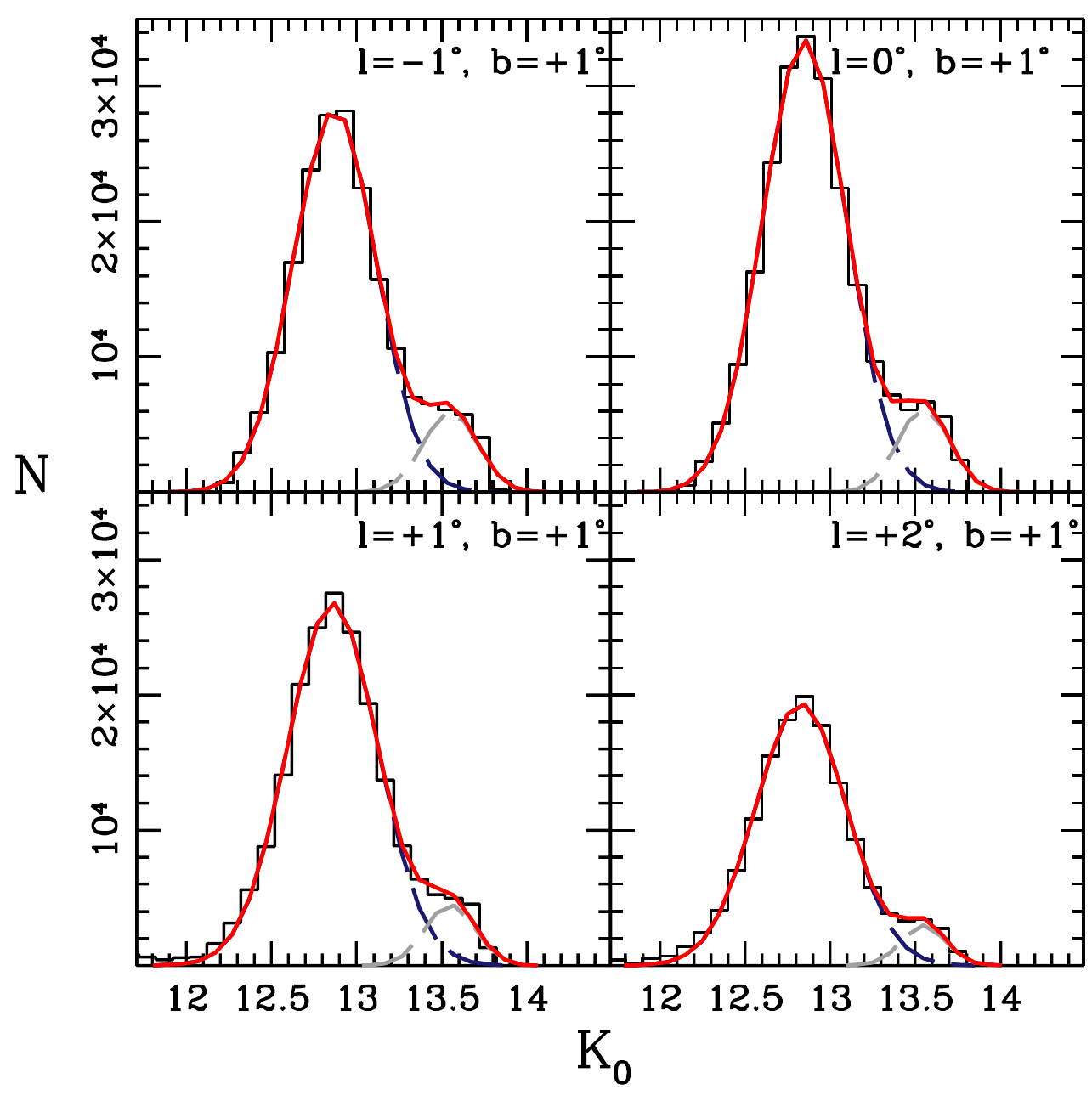}
\caption{$K_0$ LF (black histogram) for representative  fields along the Galactic plane after subtracting
the contribution from the RGB  stars. The red solid line shows the best fit to each LF, while the dashed
lines identify the individual Gaussian fits to the RC (blue) and to the additional fainter peak (grey).
\label{rcfit}}
\end{figure}

The catalogues  were corrected  for extinction applying the reddening  map of \citet{gonzalez11b} and adopting the \citet{nishiyama05} extinction law.  
From the dereddened catalogue of each field, stars  lying on the red giant branch (RGB) were selected  using a variable colour cut aimed  at excluding stars in the bright  ($K_0\le15$) main sequence  of the foreground disk,  with typical colour  $(J-K)_0\sim0.3$.  
We then used  RGB stars  to derive  the $K_0$  luminosity function  (LF), correct  it for  incompleteness, and measure the RC mean magnitude ($K^{RC}_0$) by  following the prescription of \citet{stanek97}.  
Namely, the RGB LF was fitted with a second\--order polynomial, together  with a Gaussian to account for the peak of RC stars.  
In addition, as done both by \citet{gonzalez11b} and \citet{wegg13}, a second Gaussian was included in  the  fit to  account  for  the presence  of  a  second  peak, less populated  and  fainter  than the  RC  (Fig.\,\ref{rcfit}).  
The position and relative intensity of  the second peak is broadly consistent with the expected position of  the RGB bump \citep{sweigart+90, nataf+11, wegg13}.   
However, as \cite{gonzalez11b}, we noticed that close to  the Galactic plane, the magnitude difference between the  bulge RC and this faint peak is not constant, as one would expect if it is a stellar evolutionary feature.  
The present data allowed us to map both the magnitude and  the density of this second peak with respect to the  bulge RC, and they will be the subject of a forthcoming paper.

\subsection{The RC density map}

\begin{figure*}
\centering
\includegraphics[width=15cm]{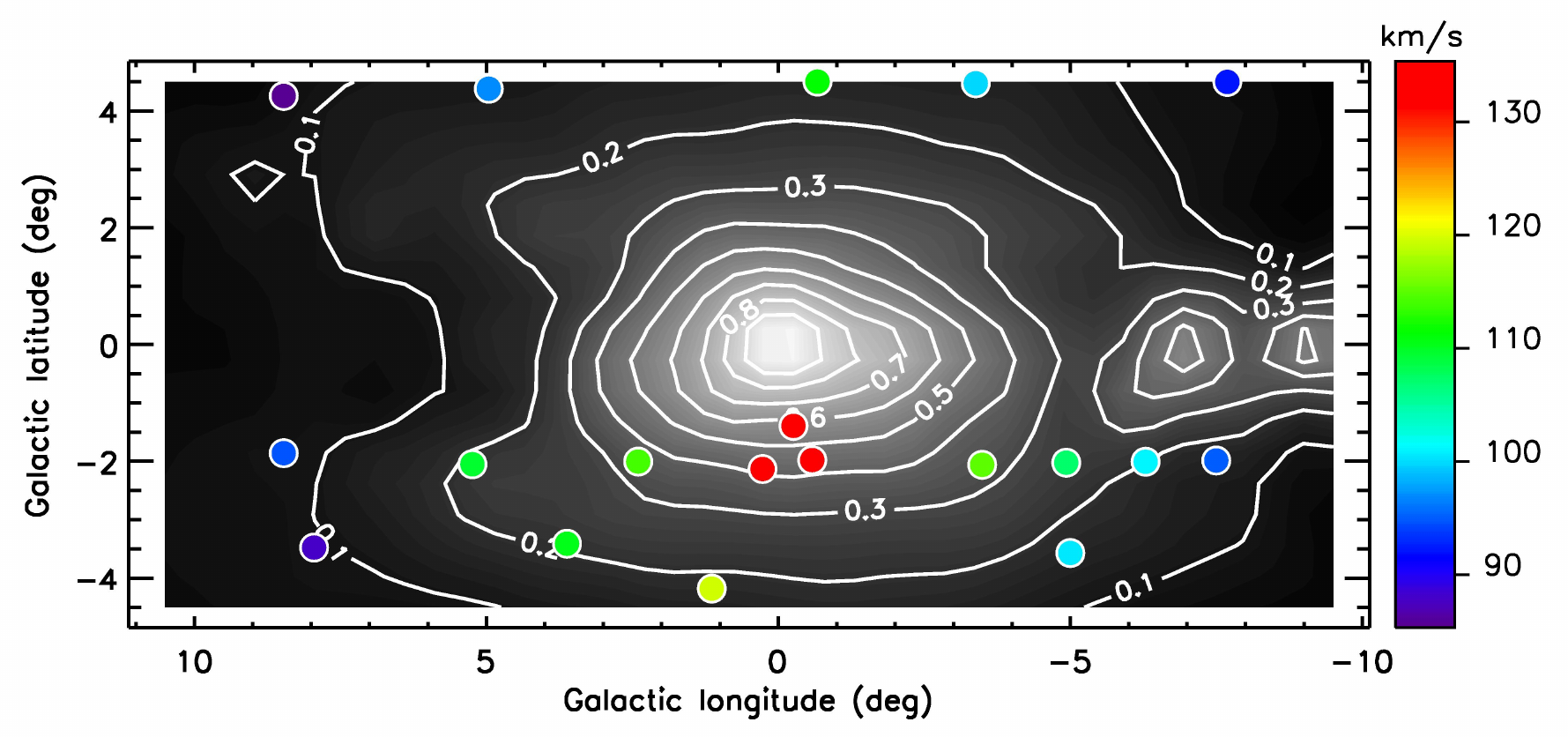}
\caption{Density map  in the  longitude-latitude plane  based on RC  star counts.  Star counts  
have  been normalised  to  the maximum  ({\it  Max}). Solid  contours are  iso-density curves  linearly spaced  by
$0.1\times$\,{\it  Max}\,deg$^{-2}$.  The fields observed by
GIBS are shown as circles,  whose colours follow the measured velocity dispersion ($\sigma$ in km/s) in that given
field. \label{densmap} }
\end{figure*}

By using the  total number of RC  stars detected in all the  fields, normalized to the  their maximum value
across the whole area, we derive the stellar density map shown in Fig.~\ref{densmap}.  
As expected, the map clearly shows  a boxy/peanut bulge, 
 with an increase  of star counts  towards the center.  
The coloured circles on top of the  grayscale map show the GIBS fields \citep{gibsI}, where radial velocities were measured for 200-400 individual RC stars per field.  
The symbols are colour\--coded according to the radial velocity dispersion measured  in each of them. 
It is clear that  the velocity dispersion follows the density of RC  stars, and hence, the mass density. 
Therefore, the present map confirms the  presence of a high stellar density  peak in the central $\sim$150\,pc,  although the peak is more  axisymmetric than what was extrapolated using the GIBS fields (see their Fig.\,11).

\begin{figure}
\centering
\includegraphics[width=8cm]{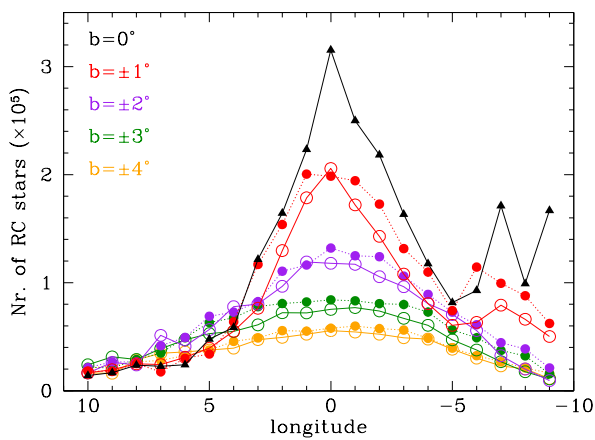}
\caption{Stellar density profile across the longitude, at fixed latitudes. Filled symbols
with dotted lines show star counts at negative latitudes, while empty symbols and
solid lines refers to positive latitudes. \label{density_profile} }
\end{figure}

The vertical extent of the iso-density contours is in general larger at positive longitude, as expected for a bar that has  its close side at positive longitude.  
This effect  was previously noted by \citet{alard01} using 2MASS photometry, although limited to $|b|\geq2^{\circ}$. 
The high quality of the VVV PSF\--fitting photometry  allows  to map  for  the  first  time the  stellar  density  of  the innermost  bulge  regions ($|b|\leq1.5^\circ$). 
The maximum density of stars is found in the region $|l|\leq 1^\circ$ and $|b|\leq 0.5^\circ$, and as shown by the 0.8 and 0.7 iso-density curves, it appears  slightly asymmetric with respect  to the bulge minor  axis.  
The central asymmetry becomes  more noticeable  when  progressively moving  outwards.  
 In fact,  as shown  in Fig.\,\ref{density_profile}, in the region between  $|l|\leq5^{\circ}, |b|\leq2^{\circ}$ we find that the counts  are systematically  higher at  $b<0$. 
 The  additional  overdensity  along the  Galactic plane  at $l\sim-7^{\circ}$  is  also present  in  the  GLIMPSE 4.5$\mu$m map  \citep{churchwell09}.
Finally, as  discussed before, the absolute  density of the central  field and of the  field at ($-7^\circ, 0^\circ$) should be taken as a lower limit because those are the only two fields for which the completeness corrections at $K_0^{RC}$ are larger than 50\% (Fig.\,\ref{compl}).

\subsection{A fully empirical estimate of the bulge stellar mass}

In what follows we present a fully empirical estimate of the stellar mass of the bulge within the VVV area.
We use the empirical Initial Mass Function (IMF) measured  by \cite{zoccali00} in a NICMOS@HST
field at ($0,-6^\circ$),  integrated to yield the mass contained  in that field, and scale it  to the whole
bulge using  the ratio of  RC stars.  The  field of view (FoV)  of NICMOS is  too small to sample evolved
stars, but a complete  LF in a larger area ($\sim8'\times 8'$; from  SOFI@NTT), including the NICMOS field,
was presented in \cite{zoccali03}.  The scaling factor  between NICMOS and SOFI ($\times$609) is simply the ratio of the  FoV areas, given that  they sample the same bulge  region.  From this scaling  factor, we can derive the mass  contained in the SOFI field.  
Finally,  to obtain the total stellar mass in the VVV area we measure the scaling factor between the number  of RGB+RC stars in the whole VVV area and in the SOFI field.

\begin{figure}
\centering
\includegraphics[width=8cm]{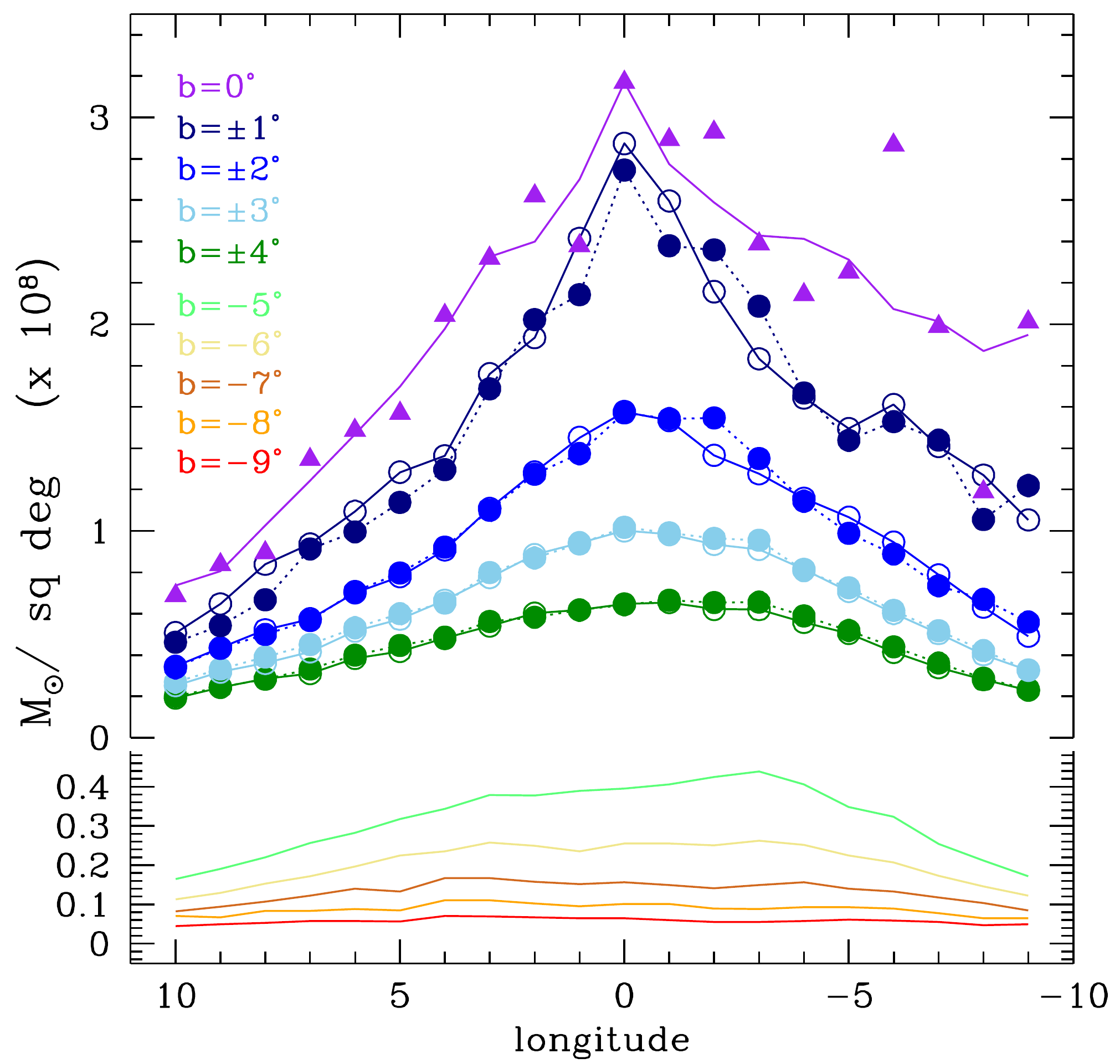}
\caption{Stellar mass profile across latitude in each field for fixed longitudes.
Filled symbols with dotted lines refer to $b<0$, empty symbols with solid lines to $b>0$ fields. The lower part of the plot has an expanded y-axis, for a better display of the high negative latitude fields. The number counts at $b=0^\circ$ are rather noisy, especially at negative longitudes, due to the higher extinction, hence higher incompleteness in those fields. The purple line includes a mild smoothing, from which the central point was excluded, in order to avoid flattening the peak. \label{mass_profile}}
\end{figure}

In practice, the LF for the VVV catalogues was restricted to the RGB at the RC level ($K_0^{RC}\pm1.5$ mag) where the completeness is high (and thus the correction small) across the whole bulge area. Because the RC magnitude varies across the fields due  to the  bar orientation, we  measured the RC  position in  each field  individually, and
counted  stars within $\pm  1.5$ mag  of the peak.  
This  procedure allowed  us to  obtain a spatial mass profile in 1$^\circ$ bins (Fig.~\ref{mass_profile}). 
 In the small region where the RC is split due to the X-shape of the bulge  ($|l|<4^\circ$, for $b<-5^\circ$) we use the mean value of the two RCs.
This is accurate  enough for our purpose, which is  just to find a normalisation factor  between the RGB in
the SOFI and in the whole  VVV area.  Furthermore, we can assume symmetry about the Galactic plane
to  extend  our  mass  estimate  to   the  $\sim$squared  region,  \-- hereafter  VVV$_{\Box}$ \--  $|l|<10^\circ$ and $|b|<9.5^\circ$ ($\sim$ 2.8$\times$2.7\,kpc, assuming a distance to the Galactic center of 8\,kpc).
  In other  words, the number of RGB+RC  stars for $b>4.5^\circ$, outside the  VVV area, is
assumed equal  to those at $b<-4.5^\circ$.   Finally, since the LF  from the SOFI field  was decontaminated
from disk stars (amounting to $18\%$ within $\pm 1.5$ mag around the RC) also the number of RGB+RC stars in VVV$_{\Box}$ was  reduced by  this quantity.  The  ratio between  the total number  of RGB+RC  stars within $\pm1.5$ mag from the RC, in VVV$_{\Box}$ vs SOFI is 58624.

In \cite{zoccali00} the IMF  was integrated using the observational data for main  sequence stars (from the
turnoff  down to  0.15  $M_\odot$) and  different  assumptions for  the IMF  above  1\,$M_\odot$ and  below
0.15\,$M_\odot$.    From  recent   determinations  of   the  IMF   of  massive   stars  \citep[see][for   a
  review]{bastian+10} we  assume a Salpeter  IMF ($\alpha=-2.35$) above 1\,$M_\odot$  and a power  law with slope  $\alpha=-0.3$  in  the substellar  regime.  With  the  same  initial-to-final mass  relation  as  in
\cite{zoccali00}, we found  570\,$M_\odot$ in the NICMOS field, hence  $570\times609\times58624 = 2.0\times
10^{10}$\,$M_{\odot}$ in VVV$_{\Box}$.   This is the mass in  stars and remnants of the  Galactic bulge in
the region $|b|<9.5^\circ$, $|l|<10^\circ$.  It assumes a  uniform disk contamination (at the level of 18\%
close to  the RC)  across the whole  bulge area.   While this  assumption is accurate  for fields  close to
($0,-6^\circ$),  it  is  likely  an overestimate  for  region  closer  to  the  Galactic  plane,  and
an underestimate for outer regions\footnote{The disk contamination  in three fields at different latitudes  was estimated in \cite{zoccali08}
  using the  Besan\c{c}on model.  At ($0,-6^\circ$) it  is very close  to $20\%$,  in Baade's Window  it is
  significantly smaller,  while at ($0,-12^\circ$) is  larger (c.f., their Table\,4).}.   
Currently, an accurate estimate of the disk  to bulge fraction for the inner few kpc of the  Galaxy is not yet available.
However by adopting  a smoothly varying  disk contamination from $20\%$ at $b=\pm9^\circ$ down  to $5\%$ at $b=\pm2^\circ$, and  then back to $20\%$  at $b=0^\circ$ where the  thin disk might become  important again \cite[e.g.,][]{dekany+15}, the total mass would increase only by $7\%$. 
Other systematic error sources, such  as the slope  of the IMF above $1\,M_\odot$ and below $0.08\,M_\odot$,
or the initial-to-final mass, are estimated as follows. A generous $\pm0.3$ dex variation of the assumed IMF slope changes the total mass by  less than $12\%$. We further assume  that small gradients on the bulge dominant stellar populations do not change the number ratio between RC and main sequence stars by more  than a  few  percent.  In  summary,  a  conservative total uncertainty  including disk  contamination, extrapolation  of the  IMF, mass  of the  remnants,  and stellar population gradients  amounts to  $15\%$, yielding a total bulge mass within $VVV_{\Box}$ of $2.0\pm 0.3\times 10^{10}$\,$M_{\odot}$.

\section{Conclusions}

We have used  the VVV $J$ and $K_s$ single\--epoch data to  derive accurate PSF\--fitting photometry in  the bulge region
between $-9.5^\circ\leq l \leq 10.5^\circ$ and $|b|  \leq 4.5^\circ$. The photometric accuracy of the catalogues, which include  
modest incompleteness corrections due to crowding and extinction, allowed  to build the first density profile  of the inner bulge towards the  Galactic plane by
means of  RC star  counts. We  found a  nice agreement between  the measured RC density  distribution and  the velocity
dispersion map provided  by the GIBS survey. In  particular, the central high $\sigma$\--peak  at $|l| \leq
2^\circ$ and  $|b| \leq 3^\circ$ in  the derived  velocity dispersion curve  is well  matched by  a high
stellar density peak in the central $\sim$150\,pc. This  tight link between the stellar density profile and
the velocity dispersion  map confirms that  the flattening of the $K^{RC}_0$  trend with longitude,
for  $|l|\leq  4^\circ$,  \citep{nishiyama05,gonzalez11c} is  more  likely  due  to  an inner  compact  and
axisymmetric spheroid rather than to the presence of a nuclear bar \citep{Gerhard12}.

We further exploit the depth  and high quality of the new VVV photometry that extends also to the innermost bulge regions and
derive  the first  purely  empirical  estimate  of  the bulge  stellar  mass  by  scaling the observed RC stellar density map 
with  the  observed bulge LF  from
\citet{zoccali00, zoccali03}. We measure the mass in star and  stellar remnants of the
Galactic bulge  to be  $2.0\pm 0.3 \times  10^{10}M_{\odot}$.  Previous bulge  mass estimates,  all model
dependent, span a  considerable range from $6.1\times10^9$ \citep{robin12}  to $1.6\times10^{10}$ $M_\odot$
\citep{sofue13,portail15}.  A possible reason why our result, although compatible within the quoted errors,
is larger than the mass estimated from dynamical models \citep[e.g.][]{portail15} is the fact that our sampled volume is not limited along the line of sight. In addition, the kinematics used so far to constrain the models do not extent to the innermost bulge regions ($|l|\leq1^\circ$ and $b=-1^\circ, -2^\circ$) where a velocity dispersion peak was found \citep{gibsI}.

\begin{acknowledgements}

The authors gratefully acknowledge  the use of data from the ESO Public  Survey program ID 179.B-2002 taken
with the VISTA telescope,  data products from the Cambridge Astronomical Survey Unit.  Support for M.Z. and
D.M.  is provided  by the  BASAL CATA  Center for  Astrophysics and  Associated Technologies  through grant
PFB-06, and the Ministry for the Economy, Development, and Tourism's Programa Iniciativa Cientifica Milenio
through grant IC120009,  awarded to Millenium Institute  of Astrophysics (MAS).  D.M.  and M.Z. acknowledge
support from FONDECYT Regular grant No.  1130196  and 1150345, respectively.  J.A.G acknowledges support by
the FIC-R Fund, allocated to the project 30321072. 
MH acknowledges financial support from the Chilean BASAL Centro de Excelencia en Astrofisica y Tecnologias Afines (CATA) grant PFB-06/2007.
\end{acknowledgements}

\bibliographystyle{aa}
\bibliography{mybiblio}

\begin{thebibliography}{29}
\expandafter\ifx\csname natexlab\endcsname\relax\def\natexlab#1{#1}\fi

\bibitem[{{Alard}(2001)}]{alard01}
{Alard}, C. 2001, \aap, 379, L44

\bibitem[{{Alonso-Garc{\'{\i}}a} {et~al.}(2015){Alonso-Garc{\'{\i}}a},
  {D{\'e}k{\'a}ny}, {Catelan}, {Contreras Ramos}, {Gran}, {Amigo}, {Leyton}, \&
  {Minniti}}]{alonsoG15}
{Alonso-Garc{\'{\i}}a}, J., {D{\'e}k{\'a}ny}, I., {Catelan}, M., {et~al.} 2015,
  \aj, 149, 99

\bibitem[{{Alonso-Garc{\'{\i}}a} {et~al.}(2012){Alonso-Garc{\'{\i}}a}, {Mateo},
  {Sen}, {Banerjee}, {Catelan}, {Minniti}, \& {von Braun}}]{javier_dophot}
{Alonso-Garc{\'{\i}}a}, J., {Mateo}, M., {Sen}, B., {et~al.} 2012, \aj, 143, 70

\bibitem[{{Bastian} {et~al.}(2010){Bastian}, {Covey}, \& {Meyer}}]{bastian+10}
{Bastian}, N., {Covey}, K.~R., \& {Meyer}, M.~R. 2010, \araa, 48, 339

\bibitem[{{Churchwell} {et~al.}(2009){Churchwell}, {Babler}, {Meade},
  {Whitney}, {Benjamin}, {Indebetouw}, {Cyganowski}, {Robitaille}, {Povich},
  {Watson}, \& {Bracker}}]{churchwell09}
{Churchwell}, E., {Babler}, B.~L., {Meade}, M.~R., {et~al.} 2009, \pasp, 121,
  213

\bibitem[{{D{\'e}k{\'a}ny} {et~al.}(2015){D{\'e}k{\'a}ny}, {Minniti},
  {Majaess}, {Zoccali}, {Hajdu}, {Alonso-Garc{\'{\i}}a}, {Catelan}, {Gieren},
  \& {Borissova}}]{dekany+15}
{D{\'e}k{\'a}ny}, I., {Minniti}, D., {Majaess}, D., {et~al.} 2015, ArXiv
  e-prints [\eprint[arXiv]{1509.08402}]

\bibitem[{{Gerhard} \& {Martinez-Valpuesta}(2012)}]{Gerhard12}
{Gerhard}, O. \& {Martinez-Valpuesta}, I. 2012, \apjl, 744, L8

\bibitem[{{Gonzalez} \& {Gadotti}(2015)}]{oscarrev15}
{Gonzalez}, O.~A. \& {Gadotti}, D.~A. 2015, ArXiv e-prints
  [\eprint[arXiv]{1503.07252}]

\bibitem[{{Gonzalez} {et~al.}(2011{\natexlab{a}}){Gonzalez}, {Rejkuba},
  {Minniti}, {Zoccali}, {Valenti}, \& {Saito}}]{gonzalez11c}
{Gonzalez}, O.~A., {Rejkuba}, M., {Minniti}, D., {et~al.} 2011{\natexlab{a}},
  \aap, 534, L14

\bibitem[{{Gonzalez} {et~al.}(2011{\natexlab{b}}){Gonzalez}, {Rejkuba},
  {Zoccali}, {Valenti}, \& {Minniti}}]{gonzalez11b}
{Gonzalez}, O.~A., {Rejkuba}, M., {Zoccali}, M., {Valenti}, E., \& {Minniti},
  D. 2011{\natexlab{b}}, \aap, 534, A3

\bibitem[{{McWilliam} \& {Zoccali}(2010)}]{mczoc10}
{McWilliam}, A. \& {Zoccali}, M. 2010, \apj, 724, 1491

\bibitem[{{Minniti} {et~al.}(2010){Minniti}, {Lucas}, {Emerson}, {Saito},
  {Hempel}, {Pietrukowicz}, {Ahumada}, {Alonso}, {Alonso-Garcia}, {Arias},
  {Bandyopadhyay}, {Barb{\'a}}, {Barbuy}, {Bedin}, {Bica}, {Borissova},
  {Bronfman}, {Carraro}, {Catelan}, {Clari{\'a}}, {Cross}, {de Grijs},
  {D{\'e}k{\'a}ny}, {Drew}, {Fari{\~n}a}, {Feinstein}, {Fern{\'a}ndez
  Laj{\'u}s}, {Gamen}, {Geisler}, {Gieren}, {Goldman}, {Gonzalez}, {Gunthardt},
  {Gurovich}, {Hambly}, {Irwin}, {Ivanov}, {Jord{\'a}n}, {Kerins}, {Kinemuchi},
  {Kurtev}, {L{\'o}pez-Corredoira}, {Maccarone}, {Masetti}, {Merlo},
  {Messineo}, {Mirabel}, {Monaco}, {Morelli}, {Padilla}, {Palma}, {Parisi},
  {Pignata}, {Rejkuba}, {Roman-Lopes}, {Sale}, {Schreiber}, {Schr{\"o}der},
  {Smith}, {}, {Soto}, {Tamura}, {Tappert}, {Thompson}, {Toledo}, {Zoccali}, \&
  {Pietrzynski}}]{minniti_vvv}
{Minniti}, D., {Lucas}, P.~W., {Emerson}, J.~P., {et~al.} 2010, \na, 15, 433

\bibitem[{{Nataf} {et~al.}(2010){Nataf}, {Udalski}, {Gould}, {Fouqu{\'e}}, \&
  {Stanek}}]{nataf10}
{Nataf}, D.~M., {Udalski}, A., {Gould}, A., {Fouqu{\'e}}, P., \& {Stanek},
  K.~Z. 2010, \apjl, 721, L28

\bibitem[{{Nataf} {et~al.}(2011){Nataf}, {Udalski}, {Gould}, \&
  {Pinsonneault}}]{nataf+11}
{Nataf}, D.~M., {Udalski}, A., {Gould}, A., \& {Pinsonneault}, M.~H. 2011,
  \apj, 730, 118

\bibitem[{{Nishiyama} {et~al.}(2005){Nishiyama}, {Nagata}, {Baba}, {Haba},
  {Kadowaki}, {Kato}, {Kurita}, {Nagashima}, {Nagayama}, {Murai}, {Nakajima},
  {Tamura}, {Nakaya}, {Sugitani}, {Naoi}, {Matsunaga}, {Tanab{\'e}},
  {Kusakabe}, \& {Sato}}]{nishiyama05}
{Nishiyama}, S., {Nagata}, T., {Baba}, D., {et~al.} 2005, \apjl, 621, L105

\bibitem[{{Portail} {et~al.}(2015){Portail}, {Wegg}, {Gerhard}, \&
  {Martinez-Valpuesta}}]{portail15}
{Portail}, M., {Wegg}, C., {Gerhard}, O., \& {Martinez-Valpuesta}, I. 2015,
  \mnras, 448, 713

\bibitem[{{Robin} {et~al.}(2012){Robin}, {Marshall}, {Schultheis}, \&
  {Reyl{\'e}}}]{robin12}
{Robin}, A.~C., {Marshall}, D.~J., {Schultheis}, M., \& {Reyl{\'e}}, C. 2012,
  \aap, 538, A106

\bibitem[{{Saito} {et~al.}(2012){Saito}, {Hempel}, {Minniti}, {Lucas},
  {Rejkuba}, {Toledo}, {Gonzalez}, {Alonso-Garc{\'{\i}}a}, {Irwin},
  {Gonzalez-Solares}, {Hodgkin}, {Lewis}, {Cross}, {Ivanov}, {Kerins},
  {Emerson}, {Soto}, {Am{\^o}res}, {Gurovich}, {D{\'e}k{\'a}ny}, {Angeloni},
  {Beamin}, {Catelan}, {Padilla}, {Zoccali}, {Pietrukowicz}, {Moni Bidin},
  {Mauro}, {Geisler}, {Folkes}, {Sale}, {Borissova}, {Kurtev}, {Ahumada},
  {Alonso}, {Adamson}, {Arias}, {Bandyopadhyay}, {Barb{\'a}}, {Barbuy},
  {Baume}, {Bedin}, {Bellini}, {Benjamin}, {Bica}, {Bonatto}, {Bronfman},
  {Carraro}, {Chen{\`e}}, {Clari{\'a}}, {Clarke}, {Contreras}, {Corvill{\'o}n},
  {de Grijs}, {Dias}, {Drew}, {Fari{\~n}a}, {Feinstein},
  {Fern{\'a}ndez-Laj{\'u}s}, {Gamen}, {Gieren}, {Goldman},
  {Gonz{\'a}lez-Fern{\'a}ndez}, {Grand}, {Gunthardt}, {Hambly}, {Hanson},
  {He{\l}miniak}, {Hoare}, {Huckvale}, {Jord{\'a}n}, {Kinemuchi}, {Longmore},
  {L{\'o}pez-Corredoira}, {Maccarone}, {Majaess}, {Mart{\'{\i}}n}, {Masetti},
  {Mennickent}, {Mirabel}, {Monaco}, {Morelli}, {Motta}, {Palma}, {Parisi},
  {Parker}, {Pe{\~n}aloza}, {Pietrzy{\'n}ski}, {Pignata}, {Popescu}, {Read},
  {Rojas}, {Roman-Lopes}, {Ruiz}, {Saviane}, {Schreiber}, {Schr{\"o}der},
  {Sharma}, {Smith}, {Sodr{\'e}}, {Stead}, {Stephens}, {Tamura}, {Tappert},
  {Thompson}, {Valenti}, {Vanzi}, {Walton}, {Weidmann}, \&
  {Zijlstra}}]{vvv_dr1}
{Saito}, R.~K., {Hempel}, M., {Minniti}, D., {et~al.} 2012, \aap, 537, A107

\bibitem[{{Saito} {et~al.}(2011){Saito}, {Zoccali}, {McWilliam}, {Minniti},
  {Gonzalez}, \& {Hill}}]{saito11}
{Saito}, R.~K., {Zoccali}, M., {McWilliam}, A., {et~al.} 2011, \aj, 142, 76

\bibitem[{{Salaris} \& {Girardi}(2002)}]{salaris02}
{Salaris}, M. \& {Girardi}, L. 2002, \mnras, 337, 332

\bibitem[{{Schechter} {et~al.}(1993){Schechter}, {Mateo}, \& {Saha}}]{dophot}
{Schechter}, P.~L., {Mateo}, M., \& {Saha}, A. 1993, \pasp, 105, 1342

\bibitem[{{Sofue}(2013)}]{sofue13}
{Sofue}, Y. 2013, \pasj, 65, 118

\bibitem[{{Stanek} {et~al.}(1997){Stanek}, {Udalski}, {Szyma{\'N}ski},
  {Ka{\L}u{\.Z}ny}, {Kubiak}, {Mateo}, \& {Krzemi{\'N}ski}}]{stanek97}
{Stanek}, K.~Z., {Udalski}, A., {Szyma{\'N}ski}, M., {et~al.} 1997, \apj, 477,
  163

\bibitem[{{Sweigart} {et~al.}(1990){Sweigart}, {Greggio}, \&
  {Renzini}}]{sweigart+90}
{Sweigart}, A.~V., {Greggio}, L., \& {Renzini}, A. 1990, \apj, 364, 527

\bibitem[{{Wegg} \& {Gerhard}(2013)}]{wegg13}
{Wegg}, C. \& {Gerhard}, O. 2013, \mnras, 435, 1874

\bibitem[{{Zoccali} {et~al.}(2000){Zoccali}, {Cassisi}, {Frogel}, {Gould},
  {Ortolani}, {Renzini}, {Rich}, \& {Stephens}}]{zoccali00}
{Zoccali}, M., {Cassisi}, S., {Frogel}, J.~A., {et~al.} 2000, \apj, 530, 418

\bibitem[{{Zoccali} {et~al.}(2014){Zoccali}, {Gonzalez}, {Vasquez}, {Hill},
  {Rejkuba}, {Valenti}, {Renzini}, {Rojas-Arriagada}, {Martinez-Valpuesta},
  {Babusiaux}, {Brown}, {Minniti}, \& {McWilliam}}]{gibsI}
{Zoccali}, M., {Gonzalez}, O.~A., {Vasquez}, S., {et~al.} 2014, \aap, 562, A66

\bibitem[{{Zoccali} {et~al.}(2008){Zoccali}, {Hill}, {Lecureur}, {Barbuy},
  {Renzini}, {Minniti}, {G{\'o}mez}, \& {Ortolani}}]{zoccali08}
{Zoccali}, M., {Hill}, V., {Lecureur}, A., {et~al.} 2008, \aap, 486, 177

\bibitem[{Zoccali {et~al.}(2003)Zoccali, Renzini, Ortolani, Greggio, Saviane,
  Cassisi, Rejkuba, Barbuy, Rich, \& Bica}]{zoccali03}
Zoccali, M., Renzini, A., Ortolani, S., {et~al.} 2003, A\&A, 399, 931

\end{thebibliography}

\end{document}